\documentclass[%
 reprint,
superscriptaddress,
 amsmath,amssymb,
]{revtex4-2}
\usepackage[english]{babel}
\usepackage[utf8]{inputenc}
\usepackage{graphicx}
\usepackage{url}
\usepackage{hyperref}

\begin{document}
\title{Interfacing picosecond and nanosecond quantum light pulses}

\author{Filip So\'snicki}
    \email{Filip.Sosnicki@fuw.edu.pl}
    \affiliation{Faculty of Physics, University of Warsaw, Pasteura 5, 02-093 Warszawa, Poland}

\author{Micha\l{} Miko\l{}ajczyk}
    \affiliation{Faculty of Physics, University of Warsaw, Pasteura 5, 02-093 Warszawa, Poland}
    
\author{Ali Golestani}
    \affiliation{Faculty of Physics, University of Warsaw, Pasteura 5, 02-093 Warszawa, Poland}
    
\author{Micha\l{} Karpi\'nski}
    \affiliation{Faculty of Physics, University of Warsaw, Pasteura 5, 02-093 Warszawa, Poland}

\date{23.11.2022} 

\maketitle
\small

\noindent Published in: Nature Photonics \textbf{17}, 761 (2023). 

\noindent DOI: \href{https://doi.org/10.1038/s41566-023-01214-z}{10.1038/s41566-023-01214-z}

\medskip

\textbf{\sffamily Light is a key information carrier, enabling worldwide high-speed data transmission through a telecommunication fibre network. This information-carrying capacity can be extended to transmitting quantum information (QI) by encoding it in single photons -- flying qubits. However, various QI-processing platforms operate at vastly different timescales. QI-processing units in atomic media, operating within nanosecond to microsecond timescales, and high-speed quantum communication, at picosecond timescales, cannot be efficiently linked due to orders of magnitude mismatch in the timescales or, correspondingly, spectral linewidths. In this work, we develop a large-aperture time lens using wide-bandwidth electro-optic phase modulation to bridge this gap. We demonstrate coherent, deterministic spectral bandwidth compression of quantum light pulses by more than two orders of magnitude with high efficiency. It will facilitate large-scale hybrid QI-processing by linking the ultrafast and quasi-continuous-wave experimental platforms, which until now, to a large extent, have been developing independently.}

Single photons are perfect candidates for transmitting quantum information between different quantum systems \cite{Gisin2007,Simon2017}. The timescales of such quantum light pulses can vary significantly depending on the platform employed. Single-photon pulse duration is linked to its spectral bandwidth via the time-bandwidth product, which places a lower limit on the spectral bandwidth required to support a pulse of a given duration \cite{Karpinski2021}. Two classes of QI-processing platforms can be distinguished based on the timescale and spectral bandwidth as a criterion \cite{Awschalom2021}. 

The first class comprises ultrafast systems, where the employed photon durations are on the order of single picoseconds, corresponding to hundreds of GHz of spectral bandwidths \cite{Weiner2011}. They are based mostly on optical nonlinearities such as three- or four-wave mixing, with the particular example of widely used spontaneous parametric down-conversion (SPDC) photon-pair sources and nonlinear optical gating \cite{Mosley2008, MacLean2018}. Additionally, they can be implemented with a high repetition rate. Thanks to similar spectral bandwidths used in the classical telecommunication, they are also compatible with the already existing high-speed telecom fibre networks~\cite{Essiambre2010}.

The latter class incorporates slower systems with nanosecond-long pulses and MHz up to single GHz-wide spectra. It primarily includes matter-based systems, such as single atoms  \cite{Saffman2010, Reiserer2014} or their ensembles \cite{Lvovsky2009, Jensen2011, Blatt2012}. It also incorporates colour-centres in diamond \cite{Doherty2013}, solid-state QI processing platforms \cite{Aharonovich2016,LagoRivera2021} and optomechanics \cite{Wallucks2020}. Such systems provide quantum memories with long storage time \cite{Duan2001,Wang2017,Kaczmarek2018} and single-photon nonlinearities \cite{Fushman2008,Tiarks2016} required to perform optical QI processing, however they are inherently slow. For example, a system with $10$~MHz spectral bandwidth is fundamentally limited to operate at most at $100$~ns repetition rate. 

Large-scale QI processing is envisioned to take advantage of system from both classes, creating a hybrid quantum network or a quantum Internet \cite{Kimble2008,Wehner2018}. A challenge arises to connect the two classes of quantum systems efficiently, to combine the QI processing capabilities of the slow systems with the photon generation and efficient communication capabilities, as well as the speed of the fast systems.

A possible solution would be staying within the slow regime, employing mostly matter-based platforms \cite{Reiserer2015,Cacciapuoti2001}. One could use known methods to generate spectrally narrow, temporally long photons \cite{Ou1999,LagoRivera2021}. However, narrow spectra and long photon pulses limit the overall QI-processing network performance due to low achievable repetition rates. In particular, such MHz-wide optical pulse spectra do not take advantage of the whole available optical bandwidth of wavelength-division multiplexing (WDM).

A different approach relies on directly connecting ultrafast and matter-based systems. However, it would result in spectral filtering of broadband photons \cite{Michelberger2015}, which is inherently lossy, again reducing the network performance. For example, reducing the single-photon bandwidth by a factor of $N$ is probabilistic with a success rate of approximately $1/N$. This probability scales exponentially as $\left(1/N\right)^M$  with the number $M$ of such interfaces.

Instead, a coherent interface, enabling low-loss shaping of quantum light pulses in time and spectrum, is required to efficiently combine the two classes of quantum devices to form a quantum network operating at a high repetition rate \cite{Lavoie2013, Karpinski2017}. Such an interface needs to redistribute the energy across the single-photon pulse, simultaneously increasing its duration and narrowing its spectrum. It needs to be realized using phase-only operations, without resorting to filtering or amplification, and with low insertion loss. It requires a combination of pulse propagation in a dispersive medium combined with time-dependent phase modulation.

\begin{figure*}
    \includegraphics[width=180mm]{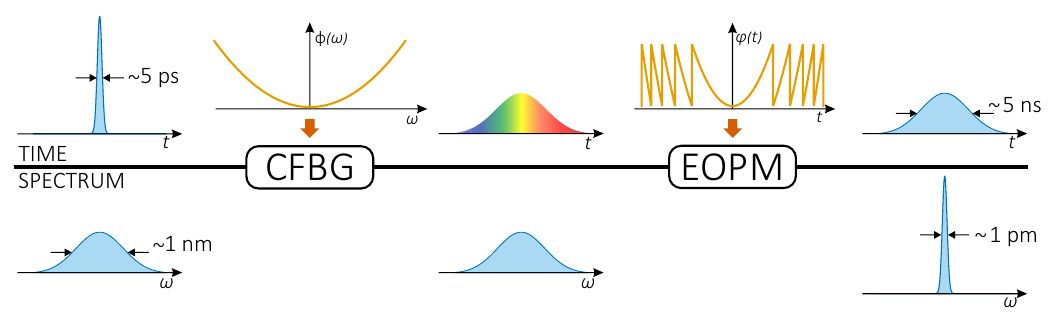}
    \caption{\textbf{Conceptual scheme of large-scale spectral bandwidth conversion}. The optical pulse's temporal profile manipulation is shown in the upper part, while its spectral manipulation is shown in the bottom. It begins with a Fourier-limited, ultrafast optical pulse, which is chirped in a highly dispersive chirped fibre Bragg grating (CFBG), increasing its duration from single picoseconds to single nanoseconds. Such chirp linearly separates different spectral components of the optical pulse in time due to applied quadratic spectral phase $\phi(\omega)$. Then time-dependent spectral shear is applied in the form of quadratic modulo 2$\pi$ temporal phase via electro-optic phase modulation (EOPM), shifting all the spectral components towards the central wavelength, hence performing a spectral compression. By using phase periodicity of 2$\pi$ one achieves ns-long temporal waveforms, resulting in compression of the spectral width of optical pulses by many orders of magnitude, to sub-GHz widths.}
    \label{fig:scheme}
\end{figure*}

\begin{figure*}
    \includegraphics[width=180mm]{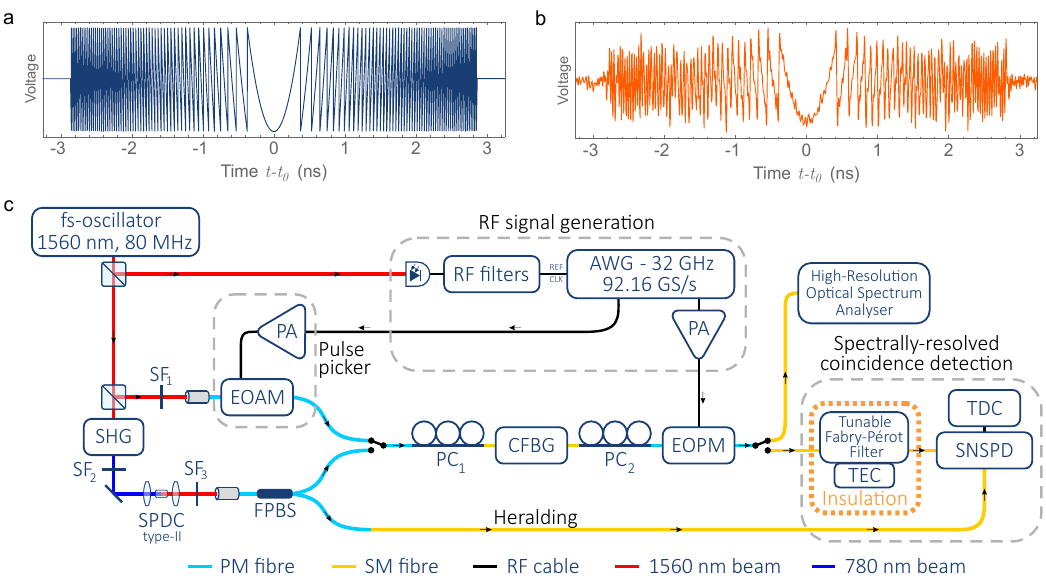}
    \caption{\textbf{Experimental setup and Fresnel waveforms.} \textbf{a} The analytical form of the Fresnel waveform $V(t)= t^{2} \,\mathrm{mod}\, 2\pi$ \textbf{b} A measured oscilloscope (63~GHz electronic bandwidth) trace of the generated Fresnel waveform, corresponding to 10~ns/nm of dispersion. It was precompensated (see Methods), transmitted through the entire RF system and measured at the RF output of the EOPM. \textbf{c} Schematic of the experimental setup. The optical pulses originating in an Er-doped femtosecond oscillator and amplifier are divided into three paths by a set of beamsplitters. In the first one, they are directly detected via a photodiode followed by a set of RF filters to generate a clock signal for radio-frequency  (RF) system synchronization. In the second path, optical pulses are spectrally shaped (SF$_1$) and passed through a home-built pulse picker using an electro-optic amplitude modulator (EOAM) to decrease the repetition rate. Then they are chirped in a chirped fibre Bragg grating (CFBG) placed between two polarization controllers (PC$_{1,2}$), spectrally compressed in an electro-optic phase modulator (EOPM) driven by an amplified (PA) RF signal from an arbitrary waveform generator (AWG) and measured by a high-resolution optical spectrum analyser. Single-photon measurements utilise the third path, where optical pulses are frequency-doubled to generate photon pairs in a type-II spontaneous parametric down-conversion (SPDC) process. The photon pairs are split in a fibre polarization beamsplitter (FPBS), where one of them -- the herald -- is sent directly to a superconducting nanowire single-photon detector (SNSPD) and time-tagged (TDC) for coincidence counting. The signal photon undergoes the same spectral compression as the classical optical pulses but is detected by another SNSPD after passing through a high-finesse Fabry-P\'erot filter. The filter is swept by a piezo element driven by a programmable power supply to retrieve information on the single-photon spectra. Additionally, the filter emulates a narrowband absorber, such as an atomic system.}
    \label{fig:setup}
\end{figure*}

The quantum interfaces based on phase-only operations demonstrated to date allow for spectral manipulations only within the ultrafast regime, i.e. from multi-THz down to tens of GHz spectral widths \cite{Allgaier2017} or only within slow regime \cite{Specht2016,Mazelanik2020}. Previous works report spectral bandwidth conversion, obtained either by optical three-wave-mixing \cite{Lavoie2013,Agha2014} and cross-phase modulation \cite{Matsuda2016}. Single-photon-level light manipulation using the four-wave-mixing scheme has also been demonstrated \cite{Salem2008,Joshi2022}. However, the nonlinear approaches suffer from high insertion loss  and optical noise due to limited conversion efficiency and the use of a strong pump. 
The electro-optic approach was the first to demonstrate efficient coherent spectral bandwidth modification of single-photon \cite{Karpinski2017, Sosnicki2020} or photon-pair pulses \cite{Mittal2017},  thanks to its deterministic nature. It has inherent unit conversion efficiency with a limiting factor of only technical losses and allows for easy central wavelength tunability. Additionally, spectral-temporal manipulation of single-photon pulses employing electro-optic phase modulation can be performed in a low-loss, all-fibre platform without the necessity of optical pumping, thus without adding optical noise to the quantum signal \cite{Karpinski2017}. However, up till now, high spectral modification factors, necessary for linking the ultrafast and slow QI processing platforms, have not been realized electro-optically due to limited electro-optic phase modulation amplitude \cite{Kolner1994,Torres2011}.

Here we show a quantum interface, based on phase-only operations, bridging the ultrafast and slow classes of QI-processing systems. We overcome the limit of electro-optic modulation amplitude by exploiting the phase periodicity, in analogy to a Fresnel lens \cite{Fresnel1822}. We replace a standard parabolic electro-optic waveform with a phase-wrapped one. By combining it with off-the-shelf low-loss dispersive elements and advanced wide-bandwidth electronics, we demonstrate a quantum interface able to efficiently compress single-photon spectral bandwidth by orders of magnitude, from picosecond to nanosecond timescales \cite{Sosnicki2018}.

Spectral bandwidth compression requires manipulation of both the temporal envelope and spectrum of an optical pulse. It consists of two stages, see Fig.~\ref{fig:scheme}. First, a spectrally wide-band optical pulse is chirped in a dispersive medium. It increases the duration of the pulse, while simultaneously linearly separating its different spectral components in time. Chirping corresponds to application of a quadratic spectral phase $ \phi(\omega) = \Phi \omega^{2}/2$, where $\Phi$ is the group delay dispersion (GDD) and $\omega$ is the angular frequency. Subsequently, a time-dependent spectral shear is applied to the pulse such that all spectral components are shifted towards a single central wavelength, reshaping the spectrum into a narrower one. It is achieved by applying quadratic temporal phase $ \varphi(t) = K t^{2}/2$, where $K$ is the chirping factor and $t$ is the retarded time. It can be also viewed as a spectral shift changing linearly in time \cite{Wright2017}. When the condition $K = \Phi^{-1}$  is met, different spectral components are shifted directly towards the centre of the pulse spectrum. One obtains it by modulating the optical pulses via electro-optic phase modulation driven by a voltage signal quadratic in time. Such an operation is called a time lens due to its mathematical analogy to a quadratic spatial phase introduced by a regular lens \cite{Kolner1994,Torres2011}.

\begin{figure}[!b]
    \centering
    \includegraphics[width = 88mm]{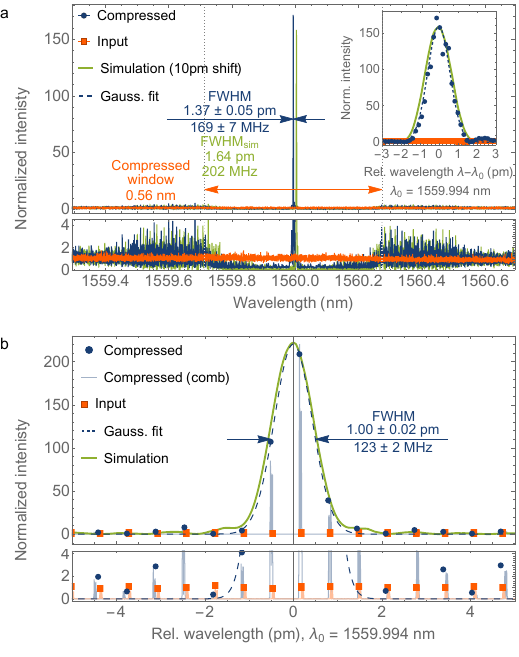}
    \caption{\textbf{Spectral compression of coherent laser pulses}. \textbf{a} Spectra of input (orange), compressed (blue) and simulated (green) coherent laser pulses using a 10~ns/nm dispersion module, normalized to the maximum of the input spectrum. The bottom sub-panel presents the same spectra for the low values of normalized intensity (also in \textbf{b}).  The inset shows the same spectrum in the narrower range of 6~pm, where each point originates from an individual longitudinal mode (see Methods), which samples the spectral envelope. To increase the number of measurement points the laser light was pulsed-picked to reduce the repetition rate to 20~MHz, hence increasing the density of longitudinal modes. The spectral compression factor of 408 and maximal spectral intensity enhancement by a factor of 154 yields compression efficiency of 40.6\% \cite{Sosnicki2018}. The plot does not include the 31.9\% transmission of the setup due to technical losses. \textbf{b} Increasing the dispersion to 15~ns/nm yields even higher enhancement of the maximal intensity. Here we disabled the pulse-picking, resulting in generation of a quasi-continuous wave (CW) light, with the output pulse duration approaching the pulse repetition rate. The light blue line shows the raw measured spectrum, with distinct longitudinal modes of the optical pulses. Due to width of the spectral envelope close to the repetition rate, which is equal to the longitudinal modes separation, only a single longitudinal mode dominates, confirming the quasi-CW regime. 
    }
    \label{fig:spectra}
\end{figure}

Chirping single-picosecond pulses to nanosecond duration requires a highly-dispersive medium with the lowest possible loss. For this reason, we use chirped fibre Bragg gratings (CFBG) with dispersion of 5~ns/nm or 10~ns/nm, commercially available at telecom wavelengths with insertion losses below 3~dB.  

The duration of the quadratic temporal phase has to match the output nanosecond optical pulse duration. The standard approach of using a single-tone RF (sine) signal driving an electro-optic phase modulator (EOPM) limits the achievable compression factor. This is because of low modulation frequency, required to cover the whole nanosecond long output optical pulse. Combined with the limit on maximal phase modulation amplitude, due to the breakdown voltage of the electro-optic modulator, it limits the achievable spectral shifts, yielding a low compression factor \cite{Kolner1994,Torres2011,Karpinski2017,Sosnicki2020}. Here we tackle this challenge by using a modulation scheme with an arbitrary temporal phase, taking advantage of the phase periodicity \cite{Sosnicki2018}, i.e.\ using a quadratic temporal phase modulo $2\pi$ factor. Such an approach, inspired by the spatial Fresnel lens \cite{Fresnel1822}, allows achieving a very long parabolic phase with steep slopes while limiting the necessary modulation amplitude to just $2\pi$, which is accessible for commercially available EOPMs. We note that related techniques has been used to generate ultrashort optical pulses from a CW laser \cite{Li2014,Fernandez2017}, however without focusing on the efficiency, crucial for quantum applications.

Generation of Fresnel-like waveforms requires using advanced  wide-bandwidth RF electronics. In particular, one needs a high-speed arbitrary waveform generator due to the waveform complexity. In contrast to the standard electro-optic time lens, where a single frequency signal is used,  a Fresnel waveform contains a wide range of frequencies, with the highest frequency components originating from the wrapping points, see Fig.~\ref{fig:setup}a. Additionally, the instantaneous frequency of a quadratic function -- its slope -- is increasing linearly in both directions from the waveform's centre, with its maximal value limited by the RF system's electronic bandwidth $f_{\mathrm{BW}}$. It induces a limitation on the waveform duration and on the maximal achievable spectral shift at the edges of the waveform \cite{Wright2017}. As a consequence, it restricts the maximal spectral width of the input optical pulses to $\Delta f = 2 f_{\mathrm{BW}}$.  Our selection of RF equipment with about 35~GHz RF bandwidth limits the maximal spectral shift to 35~GHz, which yields a 0.56~nm (70~GHz) wide spectral input window at telecom wavelengths. The non-flat frequency responses of all the RF electronics introduce distortions of the generated waveform, which contribute to the aberrations of the time lens. We used frequency response precompensation (see Methods) to counteract them. An exemplary oscilloscope (63~GHz electronic bandwidth) trace is shown in Fig.~\ref{fig:setup}b, measured at the output of the whole RF system, shown in Fig.~\ref{fig:setup}c.

First, we directly show the spectral compression employing classical light pulses using the setup shown schematically in Fig~\ref{fig:setup}c. The optical pulses from the erbium-doped fibre oscillator and amplifier were spectrally filtered (SF$_{1}$), pulse-picked, and then chirped in a CFBG module. Then a temporal phase in the form of a Fresnel waveform was applied in the EOPM. Finally, the laser pulses were detected with a high-resolution (5 MHz) optical spectrum analyser. The applied phase waveforms were generated with the AWG and synchronized to the optical pulses by its internal phase-lock loop (PLL), see Methods.

In Fig.~\ref{fig:spectra}a, we show input and compressed spectra when a chirping module with a dispersion of 10~ns/nm and a 20~MHz pulse repetition rate was used. The inset shows the same spectrum within a narrower wavelength range, where each point originates from an individual longitudinal mode of the laser, see Methods. The spectra are normalized to the maximal intensity of the input spectrum. Both spectra are measured at the output of the whole system, whose power transmission was 31.9\%. We show 154-fold (without losses), or 49-fold (including losses) enhanced maximal intensity, with full width at half maximum (FWHM) of $1.37 \pm 0.05$~pm ($169 \pm 7$~MHz), measured by fitting a Gaussian profile to the compressed spectral peak. It yields an efficiency of 40.6\%, measured as a fraction of the light intensity within the compressed peak. The non-unit value of efficiency results from the imperfections of the generated Fresnel waveform -- its aberrations, which cause shifting some spectral components in the wrong direction -- this effect contributes to low spectral peaks outside of the compressed window in Fig.~\ref{fig:spectra}a. We discuss the time lens aberrations in more detail in the supplementary section S3. The measured spectra show good agreement with simulations (see Methods) shown with a green line.

\begin{figure}[!b]
    \centering
    \includegraphics[width = 88mm]{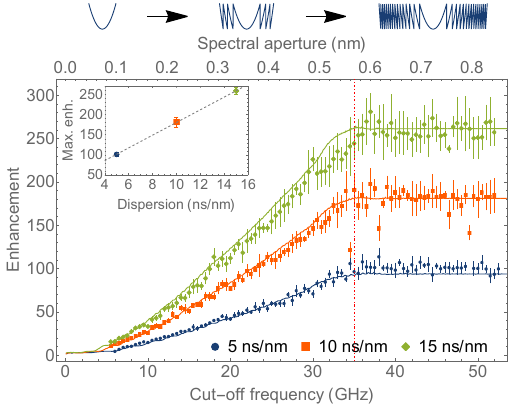}
    \caption{\textbf{Fresnel time lens aperture } By artificially limiting the Fresnel-waveform duration, i.e. the time lens aperture, as shown in the top panel, we measure the lossless enhancement for three different dispersions of 5 (blue), 10 (orange) and 15 ns/nm (green). The limitation is given by the maximal instantaneous frequency of the parabola, which also yields maximal spectral shift. The enhancement increases linearly up to the RF bandwidth $f_{\mathrm{BW}}$, shown with a red dashed line. The presented enhancement values were determined by sweeping the RF signal amplitude (in 91 steps) while measuring the spectral compression enhancement. This dependence was fitted with a Gaussian. Its maximum yielded the reported enhancement (please see Methods for more details). The error bars are based on the standard error of this Gaussian fit. Solid lines show simulation results for each value of dispersion.  The inset shows the maximal enhancement as a function of the group delay dipsersion (GDD), obtained as the mean enhancement for cut-off frequencies above the system RF bandwidth $\pm$ standard deviation. The dotted line in the inset is a linear fit. The output spectral bandwidths for these measurements are shown in supplementary figure S1.  }
    \label{fig:aper}
\end{figure}

\begin{figure}[!t]
    \centering
    \includegraphics[width = 88mm]{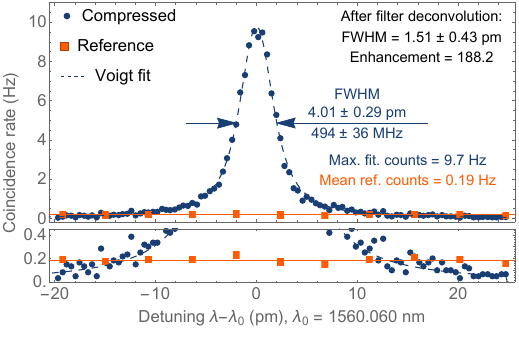}
    \caption{\textbf{Spectral manipulation of single-photon wavepackets}. Coincidence counts of photons passing through a narrowband filter with compression (blue dots) and without compression (orange squares). To avoid introducing arbitrary parameters the reference signal was measured for photons going through the same setup, but without applying temporal phase (see Methods), while the transmission of 31.9\% of the setup was measured separately. The coincidence rate was measured for 1~min/bin for compressed and 5~min/bin for reference signal. The filter was detuned to measure spectral shape of the spectrum with the results fitted with Voigt profile (see text).  The bottom sub-panel presents the same data for the low values of coincidence counts. }
    \label{fig:quantum}
\end{figure}

Next, we increased the dispersion to 15~ns/nm by joining two CFBG modules (5~ns/nm and 10~ns/nm). Additionally, we increased the repetition rate to 80~MHz, by disabling the pulse picker. By using an appropriately modified Fresnel waveform, we obtained an even higher enhancement of over 220 (without loss) and an efficiency of 38.7\%, shown in Fig.~\ref{fig:spectra}b. However, due to the use of two separate CFBG modules, introducing an increased total loss of 16\%, including the EOPM, the enhancement with system transmission taken into account was 35.2. Here we also show the generation of quasi-CW light. The FWHM of the compressed spectra is $1.00 \pm 0.02$~pm ($123 \pm 2$~MHz), which is very close to the repetition rate of 80~MHz. It means that one longitudinal mode dominates under the spectral envelope, as shown in Fig.~\ref{fig:spectra}b, where the blue, dashed line shows fitted Gaussian envelope and raw measured spectrum, with clearly visible laser longitudinal modes (comb), is shown by the light blue line (see Methods). The measured temporal profiles of the output pulses are presented in supplementary section S2 and supplementary figure S2.

To confirm the contribution of the phase-wrapped part of the Fresnel waveform to the spectral compression, we measured the enhancement when artificially limiting the duration of the waveform. Because the instantaneous frequency of the signal, proportional to its slope, increases linearly for the quadratic function, we limited the waveform up to the point where it meets the cut-off frequency $f_{\mathrm{max}}$, which we changed in the range of 5 to 52~GHz, see Fig.~\ref{fig:aper}. We show a nearly linearly increasing enhancement with the cut-off frequency up to the bandwidth of the RF system of 35~GHz, after which we reach a plateau. It clearly shows that the Fresnel-like waveforms take advantage of the whole available RF bandwidth. This experimental dataset can be further used to show the dependence of the output spectral bandwidth on the cut-off frequency, which we present in supplementary figure S1 and discuss in supplementary section S1.

To directly verify our approach to large-scale spectral compression of quantum light pulses for applications in quantum networks, we prepared pure single photons via heralded spontaneous parametric down-conversion process \cite{Mosley2008} pumped with frequency-doubled laser pulses originating from the same laser as above, see Fig.~\ref{fig:setup}c (and Methods). Then the single photons were propagated through the same 10~ns/nm CFBG and EOPM as for the case of classical light. Finally, they were detected with a superconducting nanowire single-photon detector (SNSPD) after passing through a high-finesse tunable Fabry-P\'erot interference filter with 420~MHz FWHM spectral bandwidth (see Methods), simulating a narrowband absorber like an atomic system. Then coincidences of signal and idler photons were found with a time-to-digital converter (TDC). In Fig.~\ref{fig:quantum} we show the measured 51-fold increased heralded single-photon flux of the compressed single-photon spectra (blue) in comparison to the input (orange) in the case of a lossless system. Taking into account the technical transmission of the CFBG and EOPM of 31.9\% results with the overall enhancement of a 16. It proves that such a system can be used to enhance the absorption or interference rates of bandwidth-incompatible photon pulses. The 16-fold enhancement would be further improved using a spectrally narrower absorber;  here, the absorption line was larger than the single-photon spectral width limiting the obtained enhancement. 

To find the upper bound of the enhancement for an absorber with linewidth much smaller than single-photon spectral width, we measured the single-photon spectral shape by detuning the Fabry-P\'erot interference filter. It was then fitted with the Voigt profile, shown by the blue dashed line in Fig.~\ref{fig:quantum}. Such a profile is a convolution of the single-photon spectrum's Gaussian shape and the interference filter's Lorentzian shape with a fixed width measured independently. Deconvolving the Gaussian width from the fit yields a single-photon spectral width of $1.51\pm0.43$~pm ($186 \pm 53$~MHz) and enhancement (without technical losses) of 188.2, which agrees well with the results for classical light shown in Fig~\ref{fig:spectra}a. Including the 31.9\% transmission of the system, we expect the enhancement of absorbed single photon flux by a factor of up to 60 for absorber's linewidths smaller than compressed single-photon spectral width.

We demonstrated that wideband arbitrary electro-optic temporal phase modulation enables spectral compression of classical and quantum light pulses by over 2 orders of magnitude, from tens of GHz to hundreds of MHz, reaching the regime of matter-based quantum information processing platforms. To this end we developed a Fresnel time lens, which takes the advantage of phase periodicity and all of the available RF system bandwidth. We applied this method to reduce spectral bandwidths of heralded single photons down to single pm, while enhancing their maximum in spectral intensity distribution. Additionally, delaying the RF waveform with respect to the single-photon pulses allows fine tuning the output spectrum central wavelength \mbox{\cite{Wright2017}}. By simulating a narrowband absorber with a narrowband filter we measured a 16-fold increase of photon flux through the filter as compared to direct connection. We expect to further increase this value up to 60 for absorbers with spectral linewidths narrower than the compressed pulse spectrum. Such a low-loss, all-fibre and easily reconfigurable electro-optic quantum interface will allow to increase absorption and/or multi-photon interference rates in the future hybrid quantum networks, where both ultrafast and atomic-based systems will be employed. Not only it will enable efficient network operation, but may facilitate efficient establishing of long-distance entanglement between quantum network nodes.

Development of high-speed RF devices with commercially available bandwidths of 70~GHz, as well as lower loss chirped fibre Bragg gratings will allow to further boost the performance of such an electro-optic quantum interface. Increasing the available RF bandwidth will, in particular, increase the Fresnel time lens temporal aperture, while reducing its aberrations caused by the non-ideal phase wrappings. Moreover, our device shows high potential to be used in an on-chip configuration, where bandwidths over 100~GHz with low $V_{\pi}$ have been reported in the thin-film lithium niobate platform \cite{Wang2018,Jin2021,Zhu2022,Yu2022}. Finally, employing a phase-wrapped (Fresnel) temporal phase modulation is a significant step towards arbitrary shaping of quantum light in the spectro-temporal degree of freedom \cite{Kielpinski2011,Ashby2020,Karpinski2021}.

\vspace{1em}


\vspace{4em}

\noindent\textbf{Methods}\\[1em]
{\footnotesize
\noindent\textbf{RF waveforms generation}.
The waveforms have been calculated according to the parameters: $K$ -- chirping rate of the time lens, spectral aperture $\delta f$ (see Main text), and $SR = 92.16$~GS/s being a sampling rate of the arbitrary waveform generator (AWG). The duration of the whole waveform was limited to $\Delta t = 2 \delta t = 4\pi \delta f / K$. The waveforms were then divided by the complex frequency response of the RF system (measured separately) in the frequency domain within the range of RF bandwidth of the system. The resulting waveform in the time domain was generated in the AWG (Keysight M8196A, 35~GHz RF bandwidth), amplified with a high-bandwidth power amplifier (RFLambda, 0.2 -- 35~GHz, 2~W). It was driving an electro-optic phase modulator (EOSpace, $V_{\pi@1~\mathrm{GHz}}=$~3~V, 2.4~dB insertion loss). Every change of the waveform parameters changes its minimal and maximal value. Therefore, for each new waveform, one requires to find the peak-to-peak amplitude of the RF waveform that provides the peak-to-peak phase modulation depth of $2\pi$. It was performed by sweeping the RF amplitude while measuring the spectral compression enhancement. This dependence was fitted with a Gaussian, whose centre provides the RF amplitude for $2\pi$ phase modulation depth, while its maximum yields the corresponding enhancement. The error bars in Fig.~4 are based on the standard error of this Gaussian fit.
}
\vspace{1em}

{\footnotesize
\noindent\textbf{Classical spectrum measurements}.
Spectra of classical, coherent light were acquired with a high-resolution optical spectrum analyser (APEX A2681, 5~MHz resolution). Since this resolution is higher than the repetition rate of 80~MHz or 20~MHz, individual longitudinal modes were found and summed over, creating a spectral envelope sampled with resolution given by the repetition rate. The home-built pulse picker was used to reduce the repetition rate and hence increase the resolution of measured spectra. The pulse-picker was built by utilizing a second channel of the AWG, generating rectangular pulses, which were amplified (Keysight N4985a-S50) and were driving an electro-optic amplitude modulator (Thorlabs LN05S-FC).
}
\vspace{1em}

{\footnotesize
\noindent\textbf{Source of heralded single photons}. Laser pulses originating from an erbium-doped fibre oscillator (Menlo C-Fiber 780HP) with repetition rate of 80~MHz and central wavelength 1560~nm are frequency doubled (Menlo C-Fiber 780HP) yielding a second harmonic beam with central wavelength of 780~nm. This pump beam passes through a $4f$ tunable spectral filter and is then focused into a 10~mm long periodically-polled potassium titanyl phosphate (PPKTP) crystal, where type-II spontaneous parametric down-conversion (SPDC) takes place creating, orthogonally polarized photon pairs. The pump beam is spectrally filtered such that produced photon pairs are wavelength-degenerate and spectrally uncorrelated. The photon pair beam is recollimated and additionally spectrally filtered with a set of interference filters to 1~nm FWHM such that the whole spectrum could be measured with sweeping the Fabry-P\'erot interference filter with free spectral range (FSR) of $1.2$~nm. Then the photon pairs are coupled into a polarization maintaining (PM) fibre and photons from each pair -- signal and herald (idler) -- are separated by a fibre polarisation beamsplitter (FPBS). The heralding (idler) photon is sent directly to a superconducting nanowire single-photon detector (SingleQuantum) heralding the signal photon. A time-to-digital converter (TDC, Swabian TimeTagger Ultra) is used to register coincidence events of detecting heralding idler photons and heralded signal photons.
}
\vspace{1em}

{\footnotesize
\noindent\textbf{Spectrally narrowband absorber}.
The narrowband absorber was realized with a fibre-pigtailed high-finesse tunable Fabry-P\'erot interference filter (Micron Optics FFP-TF2) with 420~MHz FWHM passband and 1.2~nm measured FSR. The tuning of the filter was automated with a stable programmable power supply (Keysight E36313A) driving a piezo-element within the filter. To thermal isolation and precise temperature control allowed to reduce the thermal drifts to below 1~pm within the measurement time. Because of a very low count rate through the filter, due to its narrowband transmission, i.e. filtering most of the photons, the coincidence count rates shown in Fig.~\ref{fig:quantum} were measured without disconnecting the CFBG and EOPM. This was to avoid introducing any arbitrary parameters such as additional polarisation mismatch from fibres reconnections or thermal drift, which could not be realigned due to the low count rate. Therefore the transmission of the setup was measured separately.
}
\vspace{1em}

{\footnotesize
\noindent\textbf{Simulations}.
The simulations were based on ref. \cite{Sosnicki2018}. Their inputs are waveforms sent to the AWG and measured input (reference) spectrum. Then the waveform is discretized in time according to the sample rate and in amplitude according to the effective number of bits (ENOB) of the AWG. Then a set of complex frequency responses of the elements is applied by multiplication. By series of applying temporal or spectral phase and fast Fourier transforms (FFT) a quadratic spectral phase and the simulated temporal phase waveform are applied to the input spectra yielding an output spectra, shown as a green line in Fig.~\ref{fig:spectra} and its enhancement shown by solid lines in Fig.\ref{fig:aper}. The simulations do not take into account a time-interval error (jitter).
}
\vspace{2em}

\noindent\textbf{Data availability}\\

The data that support the findings of this study are available from the corresponding author on reasonable request.

\vspace{2em}

\noindent\textbf{Author contributions}\\

M.K. conceived and supervised the project. F.S. designed and performed the experiment with an contribution from M.M., who built the photon-pair source and the tunable spectral filter. A. G. contributed to the early stages of the experiment. F.S and M.K. wrote the manuscript with input from all the authors. F.S. prepared the figures.

\vspace{2em}

\noindent\textbf{Acknowledgements}\\

We thank A. O. C. Davis,  M. Jachura, C. Radzewicz, B. J. Smith, N. Treps, and A. Widomski for insightful comments and discussions. We thank Keysight and AM Technologies for equipment loans. 
This work was funded in parts by the First TEAM (Project No. POIR.04.04.00-00-5E00/18) and HOMING (Project No. POIR.04.04.00-00-1E2B/16) programmes of the Foundation for Polish Science, co-financed by the European Union under the European Regional Development, and in part by the National Science Centre of Poland (Project No. 2019/32/Z/ST2/00018, QuantERA project QuICHE and Project No. 2019/35/N/ST2/04434).
\vspace{2em}

\noindent\textbf{Competing Interests}\\

The authors declare that they have no competing financial interests.

\vspace{2em}

\noindent\textbf{Correspondence}\\

Correspondence and requests for materials should be addressed to F.S. (e-mail: Filip.Sosnicki@fuw.edu.pl).

\hfill \\

\pagebreak

\onecolumngrid

\begin{center}

\textbf{\large Supplementary Information: Interfacing picosecond and nanosecond quantum light pulses}
\bigskip
\end{center}
\setcounter{equation}{0}
\setcounter{figure}{0}
\setcounter{table}{0}
\setcounter{page}{1}
\makeatletter
\renewcommand{\theequation}{S\arabic{equation}}
\renewcommand{\thefigure}{S\arabic{figure}}
\renewcommand{\thepage}{S\arabic{page}}
\renewcommand{\bibnumfmt}[1]{[S#1]}
\renewcommand{\citenumfont}[1]{S#1}

\twocolumngrid

\section{Spectral and temporal limits of the Fresnel time lens}

\begin{figure}[b]
    \centering
    \includegraphics[width = 88mm]{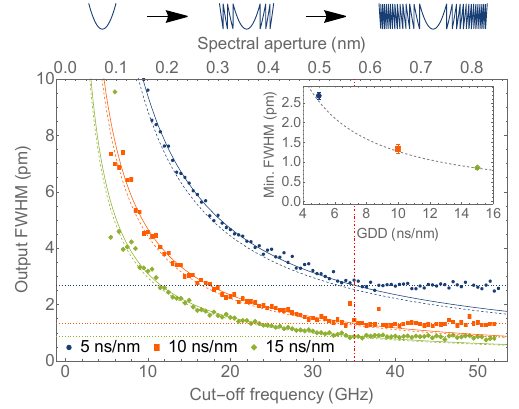}
    \caption{\textbf{Compressed spectral bandwidths}. The measured spectral bandwidths of spectrally compressed optical pulses for a range of artificially limited temporal apertures of the time lens, for three different dispersions: 5 (blue), 10 (orange), and 15~ns/nm (green), based on the same experimental dataset as Fig.~3a. The solid and dashed lines show fitted and theoretical dependencies of the output spectral widths on the RF waveform duration, respectively (see text for details). The vertical dot-dashed line marks the electronic bandwidth of the equipment. The inset shows the minimal measured output spectral bandwidth as a function of the used dispersion obtained as an average output spectral bandwidth of 34 measurement points lying to the right side of the red dot-dashed line, with error bars given by the standard deviation. The dashed line shows the theoretical dependence, assuming the time-bandwidth product of 0.886. 
    }
    \label{fig:width}
\end{figure}

We discuss the technical and fundamental limitations of the input and output spectral bandwidths in the demonstrated scheme. First, we investigate the maximal input spectral bandwidth of optical pulses that the time lens can spectrally compress, i.e., its spectral aperture. The spectral aperture has no fundamental limitation; however, it is technically limited by the RF system bandwidth, as shown in the main text when introducing Fig. 2b. It originates in the restriction of the maximal spectral shift for a given maximal RF frequency $f_{\mathrm{max}}$ and the used phase waveform's amplitude of $2\pi$, resulting in a spectral aperture of $2 f_{\mathrm{max}}$. In our demonstration, the spectral aperture is limited to about 70~GHz, due to the RF system bandwidth of about 35~GHz. Replacing the RF system with commercially available state-of-the-art components, especially the arbitrary waveform generator with the RF bandwidth of 70~GHz, would yield a maximal spectral aperture of 140~GHz. Such an approach would require employing low-$V_{\pi}$ modulators [S1] to omit the use of high-power ultra-wide-band RF amplifiers. Another possibility is to use high-order phase wrapping, i.e., by $2\pi \; k$, increasing the spectral aperture $k$-fold [S2], which will be the subject of our future research. Finally, microwave photonics enables the optical generation of ultra-wideband RF signals [S3], which in combination with low-$V_{\pi}$, high-bandwidth integrated photonics [S1] will pave the wave towards even higher spectral apertures.

To find the limits on the narrowest output spectral bandwidth, we assume that the input optical are Fourier-transform limited and are generated with the repetition rate $f_\mathrm{rep}$, such they are spaced in time with the repetition period $T_{\mathrm{rep}} = f_\mathrm{rep}^{-1}$. The duration of a Fourier-limited optical pulse $\Delta t_{\mathrm{in}}$ is linked to its spectral bandwidth $\Delta f_{\mathrm{in}}$ by the time-bandwidth product $\Delta t_{\mathrm{in}} \Delta f_{\mathrm{in}} = \mathrm{TBP}$, whose value depends on the pulse shape.

The optical pulses are then chirped in a dispersive medium with a large group delay dispersion $\Phi \gg \Delta t_{\mathrm{in}}^2$ and are modulated with a quadratic temporal phase $Kt^{2}/2$ under the condition $K=\Phi^{-1}$. In particular, the optical pulses are time-stretched by the dispersive medium, achieving frequency-to-time mapping [S4], while the applied quadratic temporal phase reorganizes the optical pulses' spectrum such that they regain the Fourier-transform limitation with a narrower spectrum. The maximal duration of the time-stretched pulses $\Delta t_{\mathrm{out}}$ is given by the repetition period $\Delta t_{\mathrm{out}} \le T_{\mathrm{rep}}$ since otherwise, the consecutive pulses will overlap, such that one cannot apply correct temporal phases onto them. In combination with the time-bandwidth product, the repetition period imposes a fundamental limit on the output spectral bandwidth as $\Delta f_{\mathrm{out}} = \frac{\mathrm{TBP}}{\Delta t_{\mathrm{out}}} \ge f_{\mathrm{rep}}$: one cannot reach a narrower spectral bandwidth than the one comparable with the repetition rate. 

In particular, saturating the above inequality constitutes a quasi-CW regime, where the duration of the optical pulses is comparable to the repetition period. Reducing the repetition rate would enable further reduction of the fundamental limit on the output spectral bandwidth. The main technical limitation on the output spectral bandwidth is due to the availability of very high group delay dispersion $\Phi$. Achieving narrower output spectral bandwidths requires chirping the same input optical pulses to a longer duration, increasing the required dispersion $\Phi$. This  dispersive medium needs to meet the technically challenging set of conditions of very high GDD, low loss, sufficient spectral bandwidth to cover the spectral aperture, and a low level of higher-order dispersion contributions. 

We experimentally show the dependence of the output spectral bandwidth on the RF waveform duration that introduces the Fresnel time lens. We artificially limit the Fresnel time lens waveform RF bandwidth. This limits the RF waveform duration and sets the maximum duration of the chirped optical signal $\Delta t_{\mathrm{out}}$  that can be bandwidth compressed. Then we measured the spectral bandwidth of the spectrally compressed light, as shown in Fig.~\ref{fig:width}. The plot shows the full-width at half maximum (FWHM) $\Delta f_{\mathrm{out}}$ of the compressed spectra for three employed dispersions of 5 (blue dots), 10 (orange squares), and 15~ns/nm (green diamonds), obtained from the same experimental data as in Fig.~3a. The solid lines show fitting functions $\frac{\mathrm{TBP}}{\Delta t_{\mathrm{out}}}$ with only one free parameter: the time-bandwidth product (TPB). Based on the fitting result, we found its values to be 0.939$\pm$0.005, 0.935$\pm$0.008, and 0.929$\pm$0.006, respectively. The uncertainty is a standard error obtained from the least squares fit. These values can be compared to the time-bandwidth product of a rectangular pulse of 0.886 since the artificially limited time lens temporal aperture selects the part of the chirped optical pulse to be compressed with hard edges. The dashed lines show the output spectral bandwidths assuming the TBP of 0.886, which is close to the experimental data. We measure the minimal spectral bandwidths for RF bandwidth of 35~GHz, marked with the vertical dotted-dashed line, which are 2.67~pm (blue dotted line), 1.34~pm (orange dotted line), and 0.86~pm (green dotted line). The inset shows the dependency of the narrowest measured FWHM of the spectrally compressed light on the GDD. The dashed line in the inset shows the theoretical function given for the TPB of 0.886.

\section{Temporal profile of output signals}

\begin{figure}[t]
    \centering
    \includegraphics[width = 88mm]{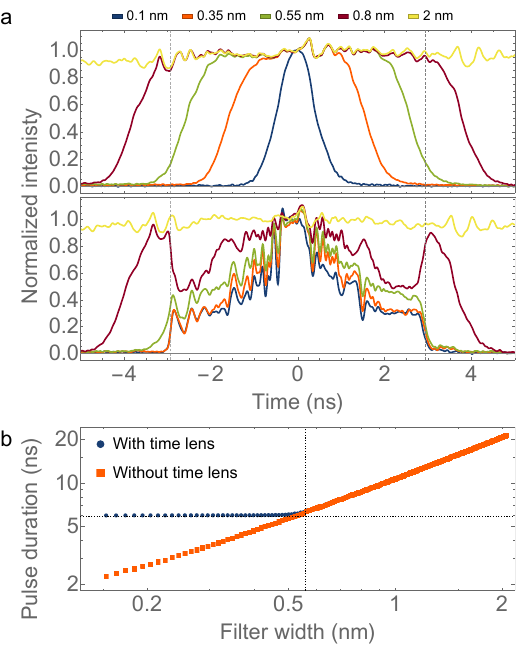}
    \caption{\textbf{Temporal measurement of spectrally compressed pulses.} \textbf{a} Oscilloscope traces of spectrally filtered optical pulses detected with high-speed photodiode without (top) and with (bottom) applying the Fresnel time lens to chirped optical pulses. Different colors show the filter bandwidth varying from 0.1~nm to 0.8~nm. The traces are normalized to the central intensity of the non-compressed pulses to neglect the spectral filter's resolution. The vertical dashed lines show the RF waveform length employing the Fresnel time lens. \textbf{b} Optical pulse duration after spectral compression, measured as the full-width at tenth maximum (FWTM), while varying the output spectral filter, with (blue dots) or without (orange squares) Fresnel time lens applied to chirped optical pulses. The vertical and horizontal dashed lines show the spectral and temporal apertures, respectively. 
    }
    \label{fig:time}
\end{figure}

The temporal profile of the output signals results from the application of the large GDD $\Phi$ to the input optical pulses. The time stretching by the GDD is accompanied by frequency-to-time mapping (chirping). The subsequent application of a quadratic temporal phase (the time lens) to the time-stretched pulses does not change its temporal envelope. We experimentally show the frequency-to-time mapping by the GDD medium by disabling the electro-optic phase modulation and placing an adjustable spectral filter at the output of the system. We measure the temporal envelope of the output classical pulses by detecting them with a high-speed photodiode (EOT ET-3500FEXT, $>$15~GHz), with the oscilloscope (Keysight DSOV134A) traces shown in the upper panel of Fig.~\ref{fig:time}a for a range of spectral bandwidths of the filter. The duration of the pulses, measured as their full-width at the tenth maximum (FWTM) for the full range of spectral bandwidths, are shown with orange squares in Fig.~\ref{fig:time}b. It shows the linearity of the frequency-to-time mapping.

Turning the time lens on -- applying the wrapped quadratic temporal phase to the optical pulses -- does not change the temporal envelope of the optical pulses leaving the time lens. The oscilloscope trace in such a case is shown with yellow lines in both subpanels of Fig.~\ref{fig:time}a when the spectral filter was fully opened (2~nm spectral bandwidth). Such a pulse contains all of the spectral components -- both compressed and non-compressed (due to the time lens aberrations and due to the limited spectral aperture). We filter out the non-compressed spectral components by narrowing the spectral filter, with the resulting oscilloscope traces from the photodiode shown in the bottom subpanel of Fig.~\ref{fig:time}a. It shows that only rough spectral filtering is necessary for removing the non-compressed parts from the optical pulses. Finally, we measure the compressed pulse FWTM duration, shown with blue dots in Fig.~\ref{fig:time}b. The minimal measured duration of the spectrally compressed pulses is 5.9~ns, which agrees with the temporal aperture of the Fresnel time lens, shown with a horizontal dashed line. The vertical dashed line shows the spectral aperture of 0.56~nm.  Further discussion of the temporal profile of the filtered output pulses is provided in the next supplementary section.

\section{Aberrations of the Fresnel time lens}

The time lens aberrations result from any deviations of the applied temporal phase from the ideal quadratic (wrapped) phase. Contrary to the standard, single-tone time lenses, where the phase features leading to aberrations are slowly varying in time [S5], the primary source of aberrations in the Fresnel time lens are the fast-changing phase-wrapping points. In particular, due to the finite bandwidth of the RF system, one cannot generate ideally sharp phase wrappings, which causes a spectral shift of some of the spectral components in the opposite direction than required to achieve spectral compression. This effect can be observed in the bottom subpanel of Fig.~3a, where one finds small spectral peaks outside the spectral aperture. Additional aberrations originate from the non-perfect RF signal precompensation, such that parts of the wrapped phase deviate from the desired quadratic phase profile, which can be seen in Fig.~2b. In particular, the RF signal amplitude varies over the duration of the RF waveform. Moreover, the precompensation may introduce high-frequency variations in the slowly-varying part of the signal around its center. In the time domain, the effects of the time lens aberrations can be observed after applying an additional spectral filter as in the previous section. Then, one filters out all of the non-compressed light, which results from the time lens aberrations. The temporal profile of the filtered output signal consists of high-frequency peaks which are approximately symmetric around the center of the pulse (bottom panel of Fig.~\ref{fig:time}a). Due to the limited RF bandwidth of measurement equipment (photodiode and oscilloscope), these oscillations are averaged out and lead to the lowered pulse intensity away from the pulse center. The time lens aberrations may be decreased by increasing the electronic bandwidth of the RF setup used to generate the RF signals.

\end{document}